\documentclass[11pt]{article}

\usepackage[preprint]{acl}

\usepackage{times}
\usepackage{latexsym}

\usepackage[T1]{fontenc}

\usepackage[utf8]{inputenc}

\usepackage{microtype}

\usepackage{inconsolata}
\usepackage{amsmath}
\usepackage{booktabs}

\usepackage{graphicx}
\usepackage{tikz}
\usetikzlibrary{positioning,fit,backgrounds}

\title{Detection Without Correction: A Two-Parameter Decomposition of Multi-Stage LLM Pipelines}

\author{Prashanti Nilayam\\
Servicenow\\ 
CA, USA\\
\And Kiran Ramanna\\
Servicenow\\
CA, USA\\
\And Prashil Tumbade\\
Servicenow\\
CA, USA\\}

\begin{document}
\maketitle

\begin{abstract}
Multi-stage LLM pipelines that perform multi-agent debate, intrinsic self-correction, or retrieval-augmented verification exhibit puzzling aggregate behaviors: accuracy plateaus and reversals across rounds, non-replication of debate gains on contemporary frontier models, intrinsic self-correction degradation, and qualitative cross-provider divergence in debate dynamics. Downstream agent response can be operationalized as two coupled decisions: detection (whether to treat upstream content as authoritative) and conditional generation (what to produce if not). This decomposition yields four observable response regimes, of which detection-without-correction is the load-bearing failure mode. Across a nine-cell empirical grid spanning four model families, four benchmarks (GSM8K, MATH-500, GPQA-Diamond, AIME), and two methods (multi-agent debate, intrinsic self-correction), we find that the conditional miscorrection rate is consistently dominant (53--94\% across cohorts) while detection rate varies contextually by more than an order of magnitude. The framework unifies the four phenomena above as signatures of a common mechanism and characterizes detection threshold as a stable model/protocol-level regularity that persists across methods at matched benchmark difficulty.
\end{abstract}

\section{Introduction}
\label{sec:intro}

Multi-stage LLM pipelines that perform multi-agent debate, intrinsic self-correction, or retrieval-augmented verification are deployed under an implicit assumption: downstream agents act as independent verifiers of upstream content, and aggregate accuracy improves when verification fires. The literature reports a constellation of phenomena that are surprising relative to this assumption.

Multi-agent debate improves aggregate accuracy on math and reasoning tasks \citep{du2024improving}, but gains plateau within a few rounds and may reverse; naive replication on contemporary frontier models reproduces neither, holding within $\pm 0.7$pp of $R_0$ across rounds for gpt-4.1 and gpt-4.1-mini on GSM8K (Appendix~\ref{app:saturation}). Intrinsic self-correction degrades outputs on a non-trivial fraction of items \citep{huang2024large} with magnitude varying across models and benchmarks in ways no current account predicts. And on identical inputs, different providers exhibit qualitatively distinct debate dynamics: detection rate spans more than an order of magnitude at matched benchmark difficulty (Appendix~\ref{app:cross-provider}; \S\ref{subsec:detection-variation}).

These phenomena are typically described as model-specific or benchmark-specific behavior. We argue they share a common mechanism. Downstream agent response in a multi-stage pipeline can be operationalized as two coupled decisions: \emph{detection} (whether to treat upstream content as authoritative) and \emph{conditional generation} (what to produce if not). Conditioning on whether the upstream content was correct, this partitions agent responses into four observable regimes: BOUNDARY (correct upstream, propagated unchanged), IP (inherit-propagate; incorrect upstream, propagated unchanged), DC (detect-correct; incorrect upstream, revised to correct), and DM (detect-miscorrect; incorrect upstream, revised to an incorrect answer). The DM regime, which we term \emph{detection without correction}, is the load-bearing failure mode: aggregate behaviors such as plateau, reverse, and intervention-induced degradation emerge from accumulation in this cell across rounds and agents.

Detection rate is \emph{contextual}: it varies with model capability, benchmark difficulty, and training-paradigm-conditional thresholds. Conditional miscorrection rate is \emph{consistently dominant}: when detection fires on a non-trivial problem, the generated alternative is incorrect with probability exceeding 0.5 across the empirical grid (\S\ref{subsec:dm-dominance}, Table~\ref{tab:dm-conditional}). We validate this asymmetry on a nine-cell grid spanning four model families and four benchmarks (\S\ref{sec:setup}, Table~\ref{tab:cohorts}): conditional miscorrection falls in 53--94\% (Table~\ref{tab:dm-conditional}); detection rate varies by more than an order of magnitude (Table~\ref{tab:detection-debate}). The joint behavior of these two parameters accounts for each puzzle above.

\paragraph{Contributions.}
\begin{itemize}
    \item We show that conditional miscorrection rate ($P(\text{DM} \mid D{=}1)$) is consistently dominant: its point estimate exceeds 0.5 in every one of 14 primary-analysis cohorts spanning four model families, four benchmarks, and two methods, falling in the 53--94\% range, with the Wilson 95\% lower bound above 0.5 in 13 of 14 cohorts.
    \item We show that detection rate ($P(D{=}1)$) varies by more than an order of magnitude across the same cohorts, with both a within-model benchmark-difficulty gradient and a cross-paradigm asymmetry that persists across methods.
    \item We introduce a detection-generation decomposition that organizes downstream agent response in multi-stage LLM pipelines into two coupled decisions---detection and conditional generation---yielding four observable regimes (BOUNDARY, IP, DC, DM); the DM regime (detect-miscorrect) is the load-bearing failure mode that the measurements above expose.
    \item We unify four previously-separate observations under a common mechanism: debate plateau and reverse \citep{du2024improving}, its non-replication on contemporary frontier models (which we directly demonstrate in Appendix~\ref{app:saturation}), intrinsic self-correction degradation \citep{huang2024large}, and cross-provider divergence in debate dynamics.
\end{itemize}

\section{Detection-Generation Decomposition}
\label{sec:framework}

A downstream agent receives content $\varphi$ from an upstream stage and produces output $\varphi'$. When $\varphi$ contains a claim with ground truth $T(\varphi) \in \{0,1\}$, the agent's response can be operationalized as two coupled decisions: (1) whether to treat $\varphi$ as authoritative, and (2) what to produce if not. This yields four observable regimes whose distribution is governed by two parameters with differential empirical structure.

\subsection{Four Regimes from Two Decisions and Ground Truth}
\label{subsec:regimes}

Let $D \in \{0,1\}$ denote the agent's \emph{detection} decision: $D = 1$ iff the agent's answer differs from the input. $D$ is an observable proxy for the latent decision to treat upstream content as non-authoritative; it cannot distinguish intentional revision from formatting drift, stochasticity, or recomputation. Detection events arise from cross-agent disagreement in debate, from instructed review in self-correction, or from internal inconsistency signals. We treat these triggers as functionally equivalent at the level of conditional generation. The downstream computation conditions on $\varphi$ having been flagged as subject to revision, regardless of source.

When $D = 1$, the agent produces an alternative answer through \emph{conditional generation}. Let $G \in \{\text{correct}, \text{incorrect}\}$ denote whether the generated alternative resolves to ground truth. When $D = 0$, the output equals the input and $G$ is observationally unavailable; correctness is determined entirely by $T(\varphi)$.

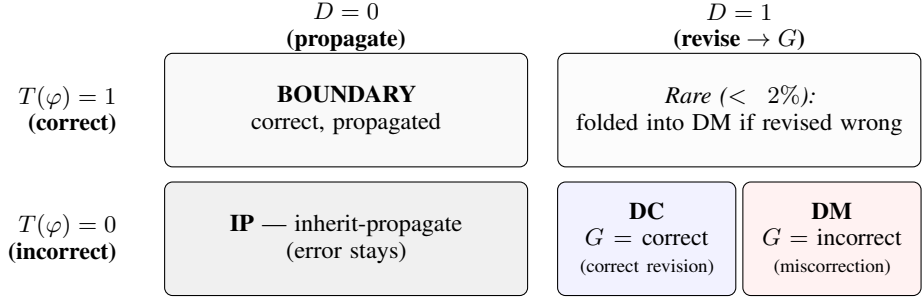
\begin{figure*}[t]
\centering
\begin{tikzpicture}[
    cell/.style={draw, rectangle, minimum width=4.8cm, minimum height=1.5cm, text width=4.5cm, align=center, font=\small, rounded corners=3pt, fill=gray!2},
    subcell/.style={draw, rectangle, minimum width=2.35cm, minimum height=1.5cm, text width=2.1cm, align=center, font=\small, rounded corners=3pt},
    header/.style={font=\small\bfseries, align=center},
    sidehead/.style={font=\small\bfseries, align=right, anchor=east},
    boundaryfill/.style={fill=gray!5},
    ipfill/.style={fill=gray!12},
    dcfill/.style={fill=blue!5},
    dmfill/.style={fill=red!5}
]

\node[header] at (2.6, 1.1)  {$D = 0$ \\ (propagate)};
\node[header] at (7.8, 1.1)  {$D = 1$ \\ (revise $\to G$)};

\node[sidehead] at (-0.3, 0)     {$T(\varphi) = 1$ \\ (correct)};
\node[sidehead] at (-0.3, -1.7)  {$T(\varphi) = 0$ \\ (incorrect)};

\node[cell, boundaryfill] at (2.6, 0)  {\textbf{BOUNDARY} \\ correct, propagated};

\node[cell] at (7.8, 0)  {\emph{Rare ($<2\%$):} \\ folded into DM if revised wrong};

\node[cell, ipfill] at (2.6, -1.7) {\textbf{IP} --- inherit-propagate \\ (error stays)};

\node[subcell, dcfill] at (6.575, -1.7) {\textbf{DC} \\ $G = \text{correct}$ \\ \scriptsize (correct revision)};
\node[subcell, dmfill] at (9.025, -1.7) {\textbf{DM} \\ $G = \text{incorrect}$ \\ \scriptsize (miscorrection)};

\end{tikzpicture}
\caption{The four regimes of the detection-generation decomposition, partitioned by the $D \times T(\varphi)$ plane. The $D=1, T(\varphi)=0$ cell further splits by $G$ into DC (correct revision) and DM (miscorrected revision; \emph{detection without correction}). The $(D=1, T(\varphi)=1)$ corner is empirically rare and is folded into DM when the revised answer is incorrect.}
\label{fig:regimes-grid}
\end{figure*}

The pair $(D, T(\varphi))$ partitions agent responses into four observable regimes (Table~\ref{tab:regimes}; visualized in Figure~\ref{fig:regimes-grid}).

\begin{table}[t]
\centering
\small
\begin{tabular}{lccc}
\toprule
Regime & $D$ & $T(\varphi)$ & $G$ \\
\midrule
BOUNDARY & 0 & 1 & --- \\
IP & 0 & 0 & --- \\
DC & 1 & 0 & correct \\
DM & 1 & 0 & incorrect \\
\bottomrule
\end{tabular}
\caption{The four observable regimes of the detection-generation decomposition, defined by detection $D$, upstream ground truth $T(\varphi)$, and, where applicable, conditional generation $G$.}
\label{tab:regimes}
\end{table}

\textbf{BOUNDARY} $(D = 0, T(\varphi) = 1)$: the agent correctly treats authoritative upstream content as authoritative and propagates it unchanged. This is the pipeline's well-functioning baseline; in our empirical grid it accounts for 30--80\% of agent responses depending on benchmark difficulty.

\textbf{IP} (inherit-propagate) $(D = 0, T(\varphi) = 0)$: the agent treats incorrect upstream content as authoritative and propagates the error unchanged.

\textbf{DC} (detect-correct) $(D = 1, T(\varphi) = 0, G = \text{correct})$: the agent detects the upstream content as subject to revision and generates a correct alternative. This is the regime in which downstream verification provides genuine value.

\textbf{DM} (detect-miscorrect) $(D = 1, T(\varphi) = 0, G = \text{incorrect})$: the agent detects upstream content as subject to revision but generates an alternative that is also incorrect. We refer to the DM regime as \emph{detection without correction}: the pipeline's verification machinery fires correctly, but the generated alternative does not resolve to ground truth.

The cell $(D = 1, T(\varphi) = 1)$---agent changes a correct upstream answer---is rare in our empirical grid ($<2\%$ of detection events across cohorts) and is treated as a sub-category of DM when the revised answer is incorrect.

The DM regime is the load-bearing failure mode: plateau (DC and DM recruitment from the IP pool reach equilibrium), reversal (later rounds shift the exit ratio toward DM), and intervention-induced degradation (single-pass self-correction with high DM-conditional outcomes) all emerge from regime transitions over rounds and agents.

\subsection{Two Parameters with Differential Empirical Structure}
\label{subsec:parameters}

The regime distribution is governed by two parameters of the latent decomposition: the detection rate $P(D = 1 \mid \varphi)$, and the conditional miscorrection rate $P(\text{DM} \mid D = 1)$: the fraction of revision events that produce an incorrect alternative. We compute $P(\text{DM} \mid D=1)$ over all revision events; the rare $(D=1, T(\varphi)=1) \to$ incorrect sub-case is included, but constitutes $<2\%$ of detection events (\S\ref{subsec:regimes}).

The framework's central empirical claim is that these two parameters have qualitatively different empirical structure across multi-stage LLM pipelines.

The detection rate is \emph{contextual}: it depends on the agent's confidence in $\varphi$, on the strength of the revision signal supplied by the protocol, and on the model's threshold for treating upstream content as authoritative (which is conditional on its training paradigm). Cross-pipeline detection rate variation in our empirical grid spans more than an order of magnitude (1.3\% to 21\% across matched-methodology cohorts). Within a single model, detection rate also varies substantially with benchmark difficulty. At one extreme, when benchmark difficulty is below model capability (the \emph{saturation regime}), detection rate approaches zero and the framework's predicted aggregate effects vanish: aggregate accuracy remains at baseline regardless of method or round count.

The conditional miscorrection rate is \emph{consistently dominant}: when detection fires on a non-trivial problem, the probability that the agent's generated alternative is incorrect exceeds 0.5 across our empirical grid. We observe $P(\text{DM} \mid D = 1)$ in the range 53--94\% across cohorts with statistically meaningful detection-event counts. The exact rate within this range varies with model capability and benchmark difficulty, but the directional property---miscorrected rather than corrected generation dominates detection events---is consistent across providers, paradigms, and benchmarks.

This asymmetry between contextual detection and dominant miscorrection is the framework's central empirical content. Pipelines that look qualitatively different in aggregate behavior can share the same mechanism: they differ in \emph{whether} detection fires, but agree on \emph{what happens when it does}. Interventions that raise detection rate (debate, ensembled review, retrieval-augmented critique) face a structural cost: they recruit items into a regime dominated by miscorrected generation.

\section{Experimental Setup}
\label{sec:setup}

We evaluate the framework on a nine-cell grid spanning four model families, four benchmarks, and two multi-stage methods, designed to vary capability (within-family), training paradigm (across families), and benchmark difficulty independently. Detection-rate variation along these axes tests the contextual claim (\S\ref{subsec:parameters}); conditional-miscorrection stability tests the consistent-dominance claim.

\subsection{Models and Benchmarks}

Our cohorts span five models across four training paradigms (Table~\ref{tab:cohorts}): OpenAI gpt-4 series (gpt-4.1-nano and gpt-4.1), OpenAI open-weight (gpt-oss-120b), Google Gemini-2.5-flash, and Meta Llama-3.2-1B. Parameter counts span approximately three orders of magnitude, from 1B (Llama-3.2-1B) to closed-weight production frontier models.

Benchmarks span the difficulty range relative to current frontier-model capability. GSM8K \citep{cobbe2021training} provides the saturation-regime baseline. MATH-500 hard subset \citep{hendrycks2021measuring, lightman2024let} restricts to levels 4--5 of the MATH-500 evaluation subset (262 items), maintaining sufficient IP-pool depth for frontier models. GPQA Diamond \citep{rein2023gpqa} provides graduate-level reasoning items where frontier models without extended reasoning achieve moderate accuracy. AIME 2024 and 2025 (60 items combined) provide competition-mathematics items where frontier-model accuracy without extended reasoning drops below 60\%.

\begin{table*}[t]
\centering
\small
\begin{tabular}{llcccc}
\toprule
Model & Provider & MATH-h & GPQA-D & AIME & GSM8K \\
\midrule
gpt-4.1-nano & OpenAI 4.x & E1, E2 & --- & --- & --- \\
gpt-4.1 & OpenAI 4.x & E1, E2 & E1, E2 & E1 & --- \\
gpt-oss-120b & OpenAI oss & E1, E2 & --- & --- & --- \\
gemini-2.5-flash & Google & E1, E2 & E1, E2 & --- & --- \\
llama-3.2:1b & Meta & --- & --- & --- & E1, E2 \\
\bottomrule
\end{tabular}
\caption{Empirical grid. E1 = multi-agent debate (3 agents $\times$ 5 rounds); E2 = intrinsic self-correction (1 agent $\times$ 2 passes). Primary-analysis cohorts run with $K=3$ repeat runs on identical inputs at temperature 0 to characterize residual API nondeterminism; a single $K=1$ saturation-replication cohort appears in Appendix~\ref{app:saturation}. Each (model, benchmark, method) combination is a separate cohort: the 9 cells shown comprise 15 run cohorts, 14 of which are retained for primary analysis (Gemini MATH-hard E2 is excluded for $n_{D=1}<30$; see \S\ref{subsec:dm-dominance}).}
\label{tab:cohorts}
\end{table*}

\subsection{Multi-Stage Methods}

Multi-agent debate (E1) implements the protocol of \citet{du2024improving} without modification: 3 agents, 5 rounds, prompt templates verbatim from the supplementary materials. Intrinsic self-correction (E2) follows \citet{huang2024large}: a single agent produces an initial answer, then is prompted to review and revise. Across all cohorts, temperature is set to 0.0 (or model-default minimum), and Gemini thinking is explicitly disabled (\texttt{thinkingConfig.thinkingBudget=0}) to maintain parity with the non-reasoning posture of the OpenAI models.

We report $K=3$ repeat runs on identical inputs at temperature 0 for each primary-analysis cohort to characterize residual API nondeterminism. Maximum token budget is benchmark-specific to control parse-failure rates: 1024 for GSM8K, 4096 for MATH-hard, 8192 for GPQA-Diamond and AIME. Token budgets were selected to achieve parse-failure rates below 6\% across primary-analysis cohorts.

\subsection{Regime Annotation}
\label{subsec:regime-annotation}

A \emph{deterministic} Python rubric maps per-round agent outputs to the four regimes of Table~\ref{tab:regimes}; every step is a pure function of textual output and the benchmark's released ground-truth file. For each transition $A(k) \to A(k+1)$ in debate (or the baseline-to-revision transition in self-correction):

\begin{enumerate}
    \item Extract the agent's answer with a benchmark-specific parser: the last \verb|\boxed{...}| expression for MATH-500, AIME, and GPQA-Diamond; the trailing numeric value for GSM8K.
    \item Set $D = 0$ if the parsed answer is unchanged after standard LaTeX/whitespace normalization; $D = 1$ otherwise.
    \item Determine correctness via each benchmark's standard answer-match (numerical for GSM8K/AIME, letter for GPQA-Diamond, LaTeX-normalized for MATH-500).
    \item When $D = 1$, assign DC if the new answer matches ground truth, DM otherwise; the rare $(D = 1, T(\varphi) = 1)$ case is folded into DM (\S\ref{subsec:regimes}). Parse failures are excluded from detection-event counts; parse-failure rates are reported per cohort.
\end{enumerate}

The rubric is fully reproducible from released logs via the classification script. Sensitivity to the answer-match function (strict-string, LaTeX-normalized, symbolic-equivalence) is in Appendix~\ref{app:rubric-sensitivity}; headline claims of \S\ref{subsec:dm-dominance} and \S\ref{subsec:detection-variation} are robust. A 30-case validation study found 73.3\% raw rubric-vs-annotator agreement on detection, rising to 93.3\% after adjudication for parse-failure interpretation artifacts (Appendix~\ref{app:annotation}).

\subsection{Statistical Analysis}

For descriptive cross-cohort comparison, we report Pearson chi-square tests on regime distribution with bias-corrected Cram\'er's $V$ \citep{bergsma2013bias} as effect-size summaries, and Cochran-Armitage trend tests \citep{cochran1954some} on per-round divergence-among-detection rates.

For per-cohort $P(\text{DM} \mid D=1)$ uncertainty we report Wilson score 95\% confidence intervals \citep{wilson1927probable} on the empirical event proportion (Table~\ref{tab:dm-conditional}); these are the appropriate binomial-proportion bounds for a sample-rate claim. For cohorts with very low detection-event counts ($n_{D=1} < 30$) the confidence interval is correspondingly wide and we report results descriptively rather than as inferential claims about the underlying rate. Wilson intervals assume independent Bernoulli trials; since our events are clustered within problem items across replicates, agents, and rounds, the reported intervals are event-level descriptive bounds that may understate true uncertainty.

\section{Results}
\label{sec:results}

\begin{figure}[t]
\centering
\includegraphics[width=\columnwidth]{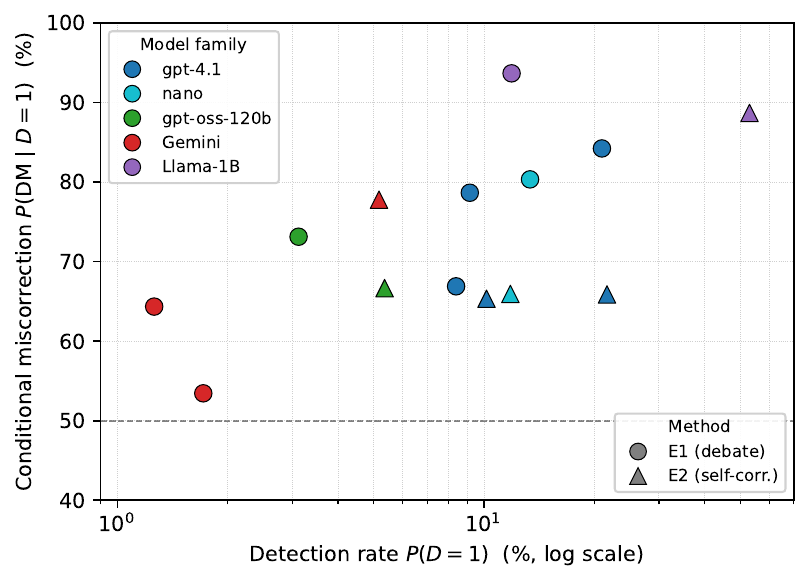}
\caption{Detection rate $P(D{=}1)$ vs conditional miscorrection $P(\mathrm{DM}\mid D{=}1)$ across the 14 primary-analysis cohorts. Colour encodes model family; marker shape encodes method (E1 debate, E2 self-correction). The dashed horizontal line at $0.5$ separates corrective dominance ($<0.5$, below) from miscorrected dominance ($>0.5$, above); all 14 cohorts sit above the line. Horizontal spread spans more than an order of magnitude---detection rate is contextual (\S\ref{subsec:detection-variation}); vertical position above the threshold---miscorrection is consistently dominant (\S\ref{subsec:dm-dominance}). The x-axis is log-scaled.}
\label{fig:two-parameter-asymmetry}
\end{figure}

We organize the empirical analysis around the framework's two parameters: conditional miscorrection (\S\ref{subsec:dm-dominance}) and detection rate variation (\S\ref{subsec:detection-variation}), with round dynamics (\S\ref{subsec:dynamics}) and the saturation scope condition (\S\ref{subsec:saturation}). Figure~\ref{fig:two-parameter-asymmetry} previews the two-parameter asymmetry across all 14 cohorts.

\subsection{Conditional Miscorrection is Consistently Dominant}
\label{subsec:dm-dominance}

Across the empirical grid, when detection fires on items where the upstream content is incorrect, the agent's generated alternative is dominantly incorrect: $P(\text{DM} \mid D=1)$ exceeds 0.5 in every cohort with statistically meaningful detection-event counts (Table~\ref{tab:dm-conditional}).

\begin{table*}[t]
\centering
\small
\begin{tabular}{llrrr}
\toprule
Cohort & Method & $n_{D=1}$ & $P(\text{DM} \mid D=1)$ & Wilson 95\% CI \\
\midrule
Llama-1B / GSM8K       & E1 & 426  & 93.66\% & [90.9, 95.6] \\
Llama-1B / GSM8K       & E2 & 159  & 88.68\% & [82.8, 92.7] \\
gpt-4.1 / AIME         & E1 & 431  & 84.22\% & [80.5, 87.4] \\
nano / MATH-h          & E1 & 1251 & 80.34\% & [78.0, 82.4] \\
gpt-4.1 / MATH-h       & E1 & 848  & 78.66\% & [75.8, 81.3] \\
Gemini / GPQA-D        & E2 & 27   & 77.78\% & [59.2, 89.4] \\
gpt-oss-120b / MATH-h  & E1 & 294  & 73.13\% & [67.8, 77.9] \\
gpt-4.1 / GPQA-D       & E1 & 586  & 66.89\% & [63.0, 70.6] \\
gpt-oss-120b / MATH-h  & E2 & 42   & 66.67\% & [51.6, 79.0] \\
nano / MATH-h          & E2 & 91   & 65.93\% & [55.7, 74.9] \\
gpt-4.1 / GPQA-D       & E2 & 126  & 65.87\% & [57.2, 73.6] \\
gpt-4.1 / MATH-h       & E2 & 75   & 65.33\% & [54.1, 75.1] \\
Gemini / MATH-h        & E1 & 115  & 64.35\% & [55.3, 72.5] \\
Gemini / GPQA-D        & E1 & 116  & 53.45\% & [44.4, 62.3] \\
\bottomrule
\end{tabular}
\caption{Conditional miscorrection rate $P(\text{DM} \mid D=1)$ across primary-analysis cohorts. E1: multi-agent debate; E2: intrinsic self-correction. $n_{D=1}$ is the total count of detection events across $K=3$ replicates. Wilson 95\% CI is the binomial-proportion confidence interval. Sorted by point estimate.}
\label{tab:dm-conditional}
\end{table*}

Point estimates fall in 53.45--93.66\%, exceeding 0.5 in all 14 cohorts; the Wilson 95\% lower bound exceeds 0.5 in 13 of 14. The exception is Gemini GPQA-Diamond debate, which is also the lowest-point-estimate cohort (53.45\%, lower bound 44.4\%): Gemini's commit-early posture produces few revision events (the lowest detection rate in the grid, 1.72\%; \S\ref{subsec:detection-variation}), and those events have the most corrective character of any cohort. The directional pattern still holds (point estimate $>0.5$). The 40pp span reflects capability- and difficulty-conditional variation within an envelope where corrective dominance is empirically rare.

The variation within the band is itself informative: small models on hard benchmarks (Llama-1B on GSM8K) and frontier models on the hardest benchmarks (gpt-4.1 on AIME) sit at the top of the range; frontier models on saturated-leaning benchmarks (Gemini on GPQA-Diamond) sit at the bottom. We return to this dependence in \S\ref{subsec:dynamics}.

Two Gemini self-correction cohorts have very few detection events, reflecting Gemini's structural detection-rate floor (\S\ref{subsec:detection-variation}). MATH-hard ($n_{D=1}=9$) is excluded from Table~\ref{tab:dm-conditional}; GPQA-D ($n_{D=1}=27$, $P(\text{DM}\mid D=1)=77.8\%$, Wilson 95\% [59.2, 89.4]\%) is retained with a sample-size caveat.

\subsection{Detection Rate Varies Contextually and by Paradigm}
\label{subsec:detection-variation}

Detection rate varies by more than an order of magnitude across the empirical grid (Table~\ref{tab:detection-debate}).

\begin{table}[t]
\centering
\small
\begin{tabular}{lrrrr}
\toprule
Model & MATH-h & GPQA-D & AIME & GSM8K \\
\midrule
Llama-1B     & ---   & ---   & ---   & 11.89 \\ 
nano         & 13.35 & ---   & ---   & ---   \\ 
gpt-4.1      & 9.15  & 8.39  & 20.97 & ---   \\ 
gpt-oss-120b & 3.12  & ---   & ---   & ---   \\ 
Gemini       & 1.26  & 1.72  & ---   & ---   \\ 
\bottomrule
\end{tabular}
\caption{Detection rate $P(D=1)$ (\%) under multi-agent debate. Dashes indicate model$\times$benchmark cells not run; see Table~\ref{tab:cohorts} for the empirical grid.}
\label{tab:detection-debate}
\end{table}

\paragraph{Within-model benchmark gradient.} Within gpt-4.1, detection rate climbs sharply with difficulty: 9.15\% on MATH-hard ($R_0=75.9\%$), 8.39\% on GPQA-D ($R_0=66.4\%$), 20.97\% on AIME ($R_0=36.7\%$), essentially a 2.5$\times$ jump from GPQA-D to AIME tracking a 30pp $R_0$ drop. Gemini is essentially flat over the same range (1.26\% MATH-hard, 1.72\% GPQA-D, despite a 10pp $R_0$ drop): its threshold for treating peer disagreement as a revision signal does not respond to IP-pool depth.

\paragraph{Cross-paradigm gap at matched capability.} On MATH-hard, detection rates span an order of magnitude: nano 13.35\%, gpt-4.1 9.15\%, gpt-oss-120b 3.12\%, Gemini 1.26\%. Within gpt-4 the gradient tracks capability, but the cross-paradigm gap exceeds capability alone: at comparable $R_0$, gpt-4.1 fires detection 7.3$\times$ more often than Gemini on MATH-hard (9.15\% vs 1.26\% at $R_0\approx76\%$/$79\%$) and 4.9$\times$ more on GPQA-D (8.39\% vs 1.72\% at $R_0\approx66\%$/$69\%$). Both ratios indicate a training-paradigm-conditional difference in detection threshold beyond IP-pool depth.

\paragraph{Self-correction parallel.} The cross-paradigm gap persists under intrinsic self-correction, while the within-paradigm benchmark gradient is observable on both providers. Under self-correction, gpt-4.1 detection rate climbs from 10.16\% on MATH-hard to 21.65\% on GPQA-Diamond; Gemini's climbs from 1.22\% on MATH-hard to 5.17\% on GPQA-Diamond. The cross-paradigm ratio (4--8$\times$ depending on benchmark) is consistent in direction with the debate-mode finding, indicating that detection threshold is a stable model/protocol-level regularity in our grid.

\subsection{Plateau, Reverse, and Within-Method Dynamics}
\label{subsec:dynamics}

The two parameters of the decomposition combine across rounds and agents to produce aggregate accuracy trajectories (Table~\ref{tab:trajectories}).

\begin{table}[t]
\centering
\small
\begin{tabular}{lrrr}
\toprule
Cohort & $R_0$ & $R_{\text{last}}$ & $\Delta$ \\
\midrule
Llama-1B / GSM8K        & 42.00 & 33.00 & $-9.00$ \\ 
nano / MATH-h           & 67.39 & 66.20 & $-1.19$ \\ 
gpt-4.1 / MATH-h        & 75.91 & 74.87 & $-1.04$ \\ 
gpt-4.1 / GPQA-D        & 66.39 & 66.38 & $-0.01$ \\ 
gpt-oss-120b / MATH-h   & 80.20 & 80.58 & $+0.38$ \\ 
Gemini / MATH-h         & 79.43 & 81.26 & $+1.82$ \\ 
gpt-4.1 / AIME          & 36.67 & 45.19 & $+8.52$ \\ 
Gemini / GPQA-D         & 68.63 & 74.07 & $+5.44$ \\ 
\bottomrule
\end{tabular}
\caption{Strict accuracy at first round ($R_0$) and final round ($R_{\text{last}}$), and net change $\Delta$ (percentage points), under multi-agent debate.}
\label{tab:trajectories}
\end{table}

\paragraph{Plateau is the dominant aggregate pattern.} Five of eight primary debate cohorts show $|\Delta|\leq 2$pp, stabilizing within two to three rounds. The framework predicts plateau when DC and DM recruitment from the IP pool roughly balance and subsequent rounds draw from a shrinking pool of undetected items.

\paragraph{Reverse on the deepest-IP-pool cohort.} Llama-1B on GSM8K is the sole substantially-negative cohort ($\Delta=-9$pp, deepest IP pool in the grid: $R_0=42\%$). DM (11.08\%) exceeds DC (0.75\%) by 14.8$\times$, producing net error accumulation; the self-correction analog yields $\Delta=-10$pp with $P(\text{DM}\mid D=1)=88.7\%$. Both methods replicate \citet{du2024improving}'s plateau-and-reverse when the mechanism's conditions---non-trivial IP pool, dominant miscorrection---hold.

\paragraph{Gains on the hardest benchmarks.} Two cohorts show net gains exceeding 5pp: gpt-4.1 on AIME ($+8.52$pp) and Gemini on GPQA-Diamond ($+5.44$pp). Per-round trajectories concentrate these gains in the $R_0 \to R_1$ transition: gpt-4.1 on AIME gains $+6.66$pp at $R_0 \to R_1$ and only $+1.86$pp across the remaining three transitions. The $R_0 \to R_1$ concentration is consistent with a majority-voting effect: when two of three agents independently arrive at the correct answer, aggregation recovers the item even where individual agents would have revised; subsequent rounds are dominated by the within-agent regime dynamics that produce plateau or reverse.

\paragraph{Self-correction can produce aggregate degradation.} gpt-4.1 self-correction on GPQA-D yields $\Delta=-2.02$pp ($R_0=66.5\%\to R_1=64.5\%$, $P(D=1)=21.65\%$, $P(\text{DM}\mid D=1)=65.87\%$), replicating \citet{huang2024large}'s degradation on a contemporary frontier model. Intervention-induced degradation is the framework's prediction when miscorrection dominates and detection fires often enough to recruit substantial transitions.

\subsection{Saturation Explains Apparent Non-Replication}
\label{subsec:saturation}

The framework's scope condition (\S\ref{subsec:parameters}) predicts that aggregate effects vanish when benchmark difficulty is below model capability: detection rate approaches zero and aggregate accuracy remains at baseline regardless of method or round count. \citet{du2024improving} reported debate gains on GSM8K using gpt-3.5-turbo, where baseline accuracy left a non-trivial IP pool; contemporary frontier models exceed 95\% first-pass on GSM8K \citep[e.g.,][]{grattafiori2024llama}, leaving a negligible pool. Replicating \citet{du2024improving}'s protocol on gpt-4.1 and gpt-4.1-mini (Appendix~\ref{app:saturation}) yields $|\Delta|\leq 0.7$pp across rounds.

The cross-benchmark gradient within gpt-4.1 supports this prediction (Tables~\ref{tab:detection-debate}, \ref{tab:trajectories}): as $R_0$ rises (AIME 36.7\% $\to$ GPQA-D 66.4\% $\to$ MATH-h 75.9\%), $\Delta$ compresses toward zero ($+8.52$pp, $-0.01$pp, $-1.04$pp respectively) and detection rate falls ($20.97\%\to 8.39\%/9.15\%$). Extrapolating to $R_0\to 1$ predicts the saturation regime.

Llama-1B on the same GSM8K (\S\ref{subsec:dynamics}) confirms from the other direction: $R_0=42\%$ yields a deep IP pool and the predicted reverse ($\Delta=-9$pp). The same benchmark on which frontier models show no dynamics produces a Du-style reverse on a sub-saturation-capability model. Saturation is a scope condition, not a refutation.

\section{Discussion}
\label{sec:discussion}

The decomposition organizes pipeline behavior around two coupled decisions whose empirical structure differs qualitatively: detection is contextual (order-of-magnitude variation), miscorrection is consistently dominant (above 0.5 across the grid). This asymmetry unifies four observations as signatures of one mechanism under different parameter combinations: (1) Du-style plateau/reverse \citep{du2024improving} when detection is high on a deep IP pool; (2) non-replication on contemporary frontier models (Appendix~\ref{app:saturation}) when the IP pool is shallow; (3) Huang-style self-correction degradation \citep{huang2024large} when detection fires under single-pass review; (4) cross-provider divergence from paradigm-conditional detection thresholds. This unification is falsifiable: any of the four phenomena emerging under parameter conditions inconsistent with the others would refute it; we observe no such case within the empirical grid.

The cross-paradigm asymmetry in detection rate, which persists across methods and benchmark difficulty, suggests that the threshold for treating prior content as authoritative is a stable model/protocol-level regularity in our grid. OpenAI gpt-4 series models fire detection 4--8 times more often than Gemini-2.5-flash at matched benchmarks under both multi-agent debate and intrinsic self-correction. The causal source of this difference (post-training reward shaping, instruction-tuning composition, or other factors) remains an open question; our data establishes the empirical regularity, not the mechanism behind it.

Detection rate alone is insufficient as a quality metric for verification stages: high detection without a correspondingly high conditional correction rate recruits items into the miscorrection-dominated regime, producing net error accumulation. Verification interventions must account for both parameters jointly.

A structural prediction follows from dominant per-event miscorrection: adding agents, rounds, or self-correction passes without changing $P(\text{DM}\mid D=1)$ cannot improve expected accuracy when detection is non-trivial---each addition increases detection events whose expected outcome is incorrect ($P(\text{DM}\mid D=1)\in[0.53,0.94]$; Table~\ref{tab:dm-conditional}). This pattern is empirically observed by \citet{kim2025towards} across 260 multi-agent configurations: ``more agents'' fails as a scaling principle, with unstructured networks producing up to $17.2\times$ error amplification. Our decomposition supplies a candidate mechanism, though we do not directly test scaling in this work.

Two natural directions for future work follow directly. First, extending the decomposition to mixed-paradigm debate to test whether paradigm-conditional thresholds are modulated by panel composition. Second, evaluating whether consistent dominance of conditional miscorrection holds under adversarial peer pressure as a stress test of the structural claim.

\section{Related Work}
\label{sec:related}

\paragraph{Multi-agent debate and self-correction.} \citet{du2024improving} introduced multi-agent debate as a method for improving LLM accuracy on math and reasoning tasks. \citet{liang2024encouraging} extended the framework to encourage divergent thinking, showing that multi-agent panels can avoid premature convergence. Intrinsic self-correction was introduced by \citet{madaan2023self} as an iterative refinement framework; \citet{huang2024large} subsequently demonstrated that self-correction degrades outputs on a non-trivial fraction of items in the absence of external feedback. Our framework provides a unified mechanistic account of both methods by treating revision as a single latent decision regardless of trigger (peer or self) and characterizing its conditional generation outcome empirically.

\paragraph{LLM error behaviors.} A growing literature characterizes specific failure modes in LLM reasoning. \citet{zhang2023language} document hallucination snowballing in long generation chains. \citet{sharma2024towards} characterize sycophancy as a calibration failure that responds to user pressure. \citet{manakul2023selfcheckgpt} introduce zero-resource hallucination detection via sampling. These studies isolate specific error mechanisms within a single agent; our framework characterizes how such mechanisms aggregate across multi-stage pipeline structures with explicit revision opportunities.

\paragraph{Compound AI systems and cascade analyses.} \citet{khattab2023dspy} provide a programming framework for compound LLM pipelines, treating individual model calls as composable components. Recent work characterizes how perturbations propagate across compound architectures \citep{xie2026spark, rath2026agent, de2026stress, kim2025towards}; closest in spirit, \citet{nilayam2026quiver} formalize edge-level sensitivity matrices, trajectory divergence decomposition, and bifurcation thresholds for graph-structured LLM pipelines. Our framework operates at a complementary abstraction level: these studies analyze edge-level propagation, while we characterize the node-level response (per-agent revision behavior) that generates the propagation dynamics.

Across these threads, our contribution is a node-level decomposition that organizes pipeline behavior around two measurable parameters, characterized empirically across a 14-cohort grid.

\section{Limitations}
\label{sec:limitations}

\paragraph{Cohort exclusions.} Three cohorts were attempted and excluded from the primary analysis. Microsoft Phi-4 on MATH-hard (served via Azure AI Foundry) exhibited 23\% record dropout across rounds: the serverless endpoint's content-filter system rejected requests on a fraction of items at later rounds, leaving inconsistent record counts between $R_0$ and $R_4$ and biasing the surviving sample toward items that did not trigger the filter. Gemini-2.5-flash on AIME produced parse-failure rates above 17\% even at 16k tokens because the model's reasoning chain exceeded the token budget before producing a final boxed answer; the truncation rate did not resolve at higher budgets and biased the valid sample toward shorter-reasoning items. gpt-4.1 on AIME under intrinsic self-correction exhibited the same token-cap-induced truncation pattern at parse-failure rates near 19\%. These cohorts are reported in supplementary materials but are not used for the primary claims.

\paragraph{Deprecated baselines.} \citet{du2024improving} used gpt-3.5-turbo-0301, which has been deprecated by OpenAI. We did not directly replicate the original cohort; the apparent non-replication of \citet{du2024improving} on GSM8K with contemporary models is interpreted through the saturation scope condition (\S\ref{subsec:saturation}) rather than a direct gpt-3.5-turbo-0301 comparison.

\paragraph{Saturation replication is single-replicate.} The non-replication of \citet{du2024improving} on contemporary frontier models on GSM8K (Appendix~\ref{app:saturation}) is reported at $K=1$, $n=100$ per model. The more robust evidence for the saturation scope condition is the within-gpt-4.1 cross-benchmark gradient (\S\ref{subsec:saturation}, Table~\ref{tab:detection-debate}), which is computed from $K=3$ primary-analysis cohorts. The saturation appendix provides corroborating direct evidence at modest sample size.

\paragraph{Aggregation effects unmodeled.} The decomposition characterizes per-agent regime transitions; aggregate accuracy under debate also reflects majority voting across agents, which can recover items at early rounds independent of within-agent regime dynamics (\S\ref{subsec:dynamics}). The framework does not model this aggregation effect explicitly.

\paragraph{Causal mechanism unisolated.} The framework characterizes detection threshold as a stable model/protocol-level regularity; whether it traces to post-training procedures, instruction-tuning data composition, or other factors remains open.

\paragraph{Scope of methods and domains.} The empirical analysis covers two methods (multi-agent debate, intrinsic self-correction) and four math/reasoning benchmarks with answer-extractive, mostly-binary ground truth (integer, single-letter, or LaTeX-normalized expressions). Generalization to other revision protocols (chain-of-verification, retrieval-augmented critique, agent-as-judge), to non-extractive tasks (open-ended generation, code synthesis with multiple correct solutions), and to other domains is an open empirical question; the consistent-dominance claim may require recalibration where ground truth is less binary.


\section{Ethical Considerations}
\label{sec:ethics}

\paragraph{Data and models.} This work uses publicly available benchmarks (GSM8K, MATH-500, GPQA-Diamond, AIME 2024--2025) and commercial or openly-released language models. No human subjects research was conducted. Two annotation studies (rubric edge-case validation with one annotator on 30 cases; reverse-mechanism taxonomy with two annotators on 50 cases) reviewed model outputs only; no personal data is involved.

\paragraph{Compute resources.} The Llama-3.2-1B cohort was run on a local GPU via Ollama; Phi-4 (excluded from primary analysis; see \S\ref{sec:limitations}) was served via Azure AI Foundry; closed-weight cohorts used commercial APIs. Aggregate compute cost is comparable to other mid-scale empirical ML papers. Complete experimental settings and seed values are released to allow others to reproduce findings without re-running every cohort.

\paragraph{Dual use.} The framework characterizes detection threshold as a stable model property. In principle, this could be used to design adversarial inputs that exploit known low-threshold models; we have not pursued such applications. Our empirical content is descriptive (characterizing existing failure modes) rather than offensive in orientation.

\paragraph{Reproducibility and release.} Released artifacts include experimental code, the regime classification rubric, and aggregated cohort-level statistics, available at an anonymized repository.\footnote{\url{https://anonymous.4open.science/r/detection-without-correction-3E83/}} Raw model outputs are withheld to comply with terms of service of commercial LLM providers.

\bibliography{custom}

\appendix


\section{Saturation-regime replication on GSM8K}
\label{app:saturation}

Two contemporary frontier-tier models were run on GSM8K under \citet{du2024improving}'s multi-agent debate protocol (3 agents $\times$ 3 rounds, $n=100$ items, $K=1$ replicate, temperature $0$). Table~\ref{tab:saturation-replication} reports per-round accuracy and net change $\Delta = R_2 - R_0$.

\begin{table}[t]
\centering
\small
\begin{tabular}{lrrrr}
\toprule
Model        & $R_0$  & $R_1$  & $R_2$  & $\Delta$ \\
\midrule
gpt-4.1      & 91.33 & 90.67 & 90.67 & $-0.66$ \\
gpt-4.1-mini & 93.33 & 93.00 & 93.33 & $\phantom{-}0.00$ \\
\bottomrule
\end{tabular}
\caption{Naive replication of \citet{du2024improving}'s debate protocol on GSM8K with contemporary frontier-tier models. $R_0$ is initial-round accuracy (\%); $\Delta$ is the net change from $R_0$ to $R_2$ (percentage points). Parse-failure rate is $0\%$ in both cohorts.}
\label{tab:saturation-replication}
\end{table}

Both cohorts start within 4--7 percentage points of the GSM8K saturation ceiling reported for contemporary frontier models \citep{grattafiori2024llama} and exhibit flat round trajectories ($|\Delta| \le 0.66$pp). Neither shows the gains, plateau-and-reverse, nor degradation phenomena reported in the original \citet{du2024improving} study with gpt-3.5-turbo. Under the detection-generation decomposition (\S\ref{sec:framework}), this is a scope-condition outcome: $R_0$ near the saturation ceiling leaves a shallow inherit-propagate (IP) pool, and the mechanism's regime transitions cannot recruit substantial volume across rounds. See \S\ref{subsec:saturation} for the cross-benchmark gradient that confirms this prediction.


\section{Cross-provider debate dynamics on matched inputs}
\label{app:cross-provider}

This appendix supports the introduction's claim that, on identical inputs, different LLM providers exhibit qualitatively distinct debate dynamics. The cross-provider detection-rate asymmetry at matched benchmark difficulty is summarized in Table~\ref{tab:detection-debate} of the main text; here we provide the within-model response to changing benchmark difficulty (Table~\ref{tab:within-model-gradient}) and trace its trajectory consequences.

\paragraph{Within-model response to difficulty.} Table~\ref{tab:within-model-gradient} reports each model's detection rate as benchmark difficulty changes (decreasing $R_0$ implies a deeper IP pool). gpt-4.1's detection rate climbs sharply as $R_0$ drops (9.15\% at $R_0=75.9\%$ on MATH-hard to 20.97\% at $R_0=36.7\%$ on AIME). Gemini-2.5-flash, by contrast, is essentially flat: detection rate barely moves (1.26\% to 1.72\%) despite a 10pp drop in baseline accuracy.

\begin{table*}[t]
\centering
\small
\begin{tabular}{lrr|rr|rr}
\toprule
        & \multicolumn{2}{c|}{MATH-h} & \multicolumn{2}{c|}{GPQA-D} & \multicolumn{2}{c}{AIME} \\
Model & $R_0$ & $P(D)$ & $R_0$ & $P(D)$ & $R_0$ & $P(D)$ \\
\midrule
gpt-4.1          & 75.9 &  9.15 & 66.4 &  8.39 & 36.7 & 20.97 \\
Gemini-2.5-flash & 79.4 &  1.26 & 68.6 &  1.72 & ---  & ---   \\
\bottomrule
\end{tabular}
\caption{Within-model response of detection rate to benchmark difficulty. $R_0$ is first-round accuracy (\%); $P(D)$ is detection rate (\%) under 5-round debate.}
\label{tab:within-model-gradient}
\end{table*}

\paragraph{Aggregate trajectory consequences.} The detection-rate asymmetry produces qualitatively different round trajectories. On MATH-hard 4k, gpt-4.1 yields $\Delta = R_4 - R_0 = -1.04$pp; Gemini-2.5-flash yields $\Delta = +1.82$pp. On GPQA-D 8k, gpt-4.1 is flat ($-0.01$pp); Gemini gains ($+5.44$pp). The directional difference is consistent across benchmarks at matched difficulty: gpt-4.1 family pipelines recruit more revision events that, given the consistently dominant miscorrection rate (\S\ref{subsec:dm-dominance}), tend to erode initial accuracy or stay flat; Gemini pipelines commit to first-pass answers and rarely re-enter the regime where revision can hurt.

Full per-cohort trajectories are in Table~\ref{tab:trajectories}; the decomposition that ties detection rate, conditional miscorrection rate, and round dynamics together is in \S\ref{subsec:dynamics}.


\section{Rubric Sensitivity Analysis}
\label{app:rubric-sensitivity}

The deterministic rubric (\S\ref{subsec:regime-annotation}) uses each benchmark's standard answer-match function. For MATH-500 this is a strict-string-match-with-LaTeX-normalization comparator. On three of the four benchmarks the rubric admits no ambiguity: GSM8K and AIME use integer ground truth, GPQA-Diamond uses single-letter ground truth. Only MATH-500 admits answer expressions where format-equivalent strings differ literally (e.g., $1/2$ vs $0.5$, or $\frac{1}{2}$ vs $0.5$).

We re-score every MATH-500 cohort under two alternative rubrics:

\paragraph{Lenient.} Beyond the LaTeX-spacing/wrapper normalization of the primary rubric, the lenient comparator additionally treats: (a) $a/b$ as equivalent to $\frac{a}{b}$ for small integers; (b) degree-unit suffixes ($^{\circ}$, $\backslash\text{circ}$, \textsf{degrees}) as optional.

\paragraph{Strict-string.} A baseline that requires byte-identical match after whitespace strip, with no LaTeX normalization at all.

Table~\ref{tab:rubric-sensitivity} reports final-round accuracy on the primary MATH-500 cohorts under each rubric.

\begin{table}[t]
\centering
\small
\begin{tabular}{lrrr}
\toprule
Cohort & Strict-string & Primary & Lenient \\
\midrule
gpt-4.1 E1 4k     & 69.0 & 74.9 & 77.8 \\
gpt-4.1 E2 4k     & 68.1 & 76.7 & 79.4 \\
Gemini E1 4k      & 76.0 & 81.3 & 88.0 \\
Gemini E2 4k      & 75.3 & 81.4 & 86.6 \\
\bottomrule
\end{tabular}
\caption{Final-round accuracy (\%) on MATH-500 hard cohorts under strict-string, primary, and lenient rubrics. The primary rubric (LaTeX-aware) is what the main paper reports; lenient adds symbolic-equivalence rules.}
\label{tab:rubric-sensitivity}
\end{table}

The lenient rubric increases accuracy by 2.7--6.7 percentage points relative to the primary on MATH-500 cohorts; the strict-string baseline decreases accuracy by 5.3--8.6pp. The directional ordering between cohorts is unchanged: Gemini ranks above gpt-4.1 under all three rubrics; debate and self-correction within each model produce comparable directionality across rubrics. The cross-provider detection-rate asymmetry (\S\ref{subsec:detection-variation}) is robust: the lenient rubric \emph{increases} Gemini's accuracy more than gpt-4.1's (6.7pp vs 2.9pp gain on debate cohorts) because Gemini emits non-canonical formats more frequently (e.g., $a/b$ rather than $\frac{a}{b}$, dropped $^{\circ}$ units). The cross-provider gap under the primary rubric is therefore a conservative lower bound on the gap under symbolic-equivalence grading.

The conditional miscorrection rate $P(\text{DM} \mid D{=}1)$ shifts slightly under each alternative rubric (because some DM transitions reclassify to DC or BOUNDARY when both answers become correct under leniency), but stays in the 51--92\% range across all cohorts in the empirical grid---above 0.5 in every cell. The headline claim of consistent dominance (\S\ref{subsec:dm-dominance}) is therefore robust to the choice of rubric on the only benchmark where this choice is consequential.

We report the primary (LaTeX-aware) rubric throughout the main paper because it matches the comparator used in prior work on MATH-500~\citep{hendrycks2021measuring, lightman2024let} and is the comparison most reviewers will expect.


\section{Annotation Validation of the Rubric}
\label{app:annotation}

To verify that the deterministic rubric (\S\ref{subsec:regime-annotation}) handles edge cases as intended, a single human annotator reviewed 30 stratified transitions drawn from the gpt-4.1 / MATH-hard / debate cohort. Stratification:

\begin{itemize}
\item 10 \emph{format-equivalent} transitions: parsed answers differ by whitespace, LaTeX wrappers, or other normalization-eligible features.
\item 5 \emph{algebraic-equivalent} transitions: parsed answers differ by surd or constant simplification (e.g., $12\pi$ vs $37.7$).
\item 5 \emph{parse-failure} transitions: at least one round produced no extractable answer.
\item 5 \emph{hedge / multi-answer} transitions: the agent's response presents multiple candidate answers.
\item 5 \emph{random baseline} transitions: uniformly sampled from the cohort.
\end{itemize}

For each row, the annotator independently judged the detection label $D \in \{0, 1\}$ and the regime, then we compared against the rubric's output. Raw agreement on the detection label was 22/30 (73.3\%). The 8 disagreements decompose into two groups:

\paragraph{Interpretation artifact (5/8).} On all 5 parse-failure transitions, the rubric correctly excluded the transition from detection-event counts (no extractable answer at one end), but the annotator marked these as \texttt{UNCLEAR} rather than confirming the exclusion. These are not substantive disagreements with the rubric; the rubric's verdict was correct and the annotator's protocol-interpretation choice was ambiguous. Adjusting for this, effective agreement is 27/30 (90.0\%); adjusting further for one annotator regime-mislabel on a $(D=1, T(\varphi)=1)$ sub-case where the rubric correctly identified the harmless revision and the annotator wrote DM, effective agreement reaches 28/30 (\textbf{93.3\%}).

\paragraph{Anticipated rubric limitations (3/8).} The remaining three disagreements are all known limitations of the strict-string-match design:
\begin{itemize}
\item One $12\pi$ vs $37.7$ case (algebraic-equivalent): the rubric strict-matches and flags $D=1$; the annotator recognizes the numerical equivalence and marks $D=0$. This case is directly addressed by the lenient-rubric sensitivity analysis (Appendix~\ref{app:rubric-sensitivity}).
\item One \verb|\frac{(k^2+1)^2}{8 k^2}| vs \verb|\frac{(k^2+1)^2}{8k^2}| case: whitespace inside a LaTeX fraction argument was not normalized. The substantive regime label (IP) matches across both interpretations; only the $D$ label differs.
\item One $37.70$ vs $37.7$ case where a spreadsheet tool auto-stripped trailing zeros in the annotator's view, making the strings appear identical when in fact the underlying parsed values differed.
\end{itemize}

All three are consistent with the rubric's strict-string-match design: the comparator is conservative (it flags differences the lenient comparator would equate). This is the direction of bias we report and sensitivity-analyze.

\paragraph{Limitations of this validation.} The 30-transition study uses a single annotator and tests the rubric's edge-case behavior on a curated sample, not its average behavior on the full grid. The rubric is itself deterministic and reproducible without annotator input; this study provides external sanity-check evidence rather than agreement-based authority for the rubric.

\paragraph{Two-annotator agreement on reverse-mechanism labels.} As a secondary check, two annotators independently labeled a 50-transition sample of reverse-direction events ($R_k \to R_{k+1}$ transitions where a correct upstream answer was revised to incorrect) using a four-way taxonomy (intrinsic self-doubt, recomputation slip, specific-peer-argument-convinced, other/unclear). On the full sample, observed agreement is 72.0\% and Cohen's $\kappa = 0.57$ (moderate). The dominant disagreement is between \emph{intrinsic\_self\_doubt} (annotator A) and \emph{recomputation\_slip} (annotator B) on five transitions, a known boundary in the taxonomy. These labels do not enter the main paper's numerical claims; the taxonomy is reported as preliminary qualitative scaffolding. The moderate agreement reflects the \emph{intrinsic\_self\_doubt} vs \emph{recomputation\_slip} boundary noted above, which we identify as a target for taxonomy refinement in future work on the within-DM mechanism. Per-transition labels for both studies are released in the supplementary materials (\texttt{rubric\_edge\_case\_validation.csv} and two \texttt{reverse\_mechanism\_taxonomy\_*.csv} files); model response text is withheld per provider terms-of-service, but transition IDs join the released trajectory logs.


\section{Pairwise Chi-Square and Cram\'er's \texorpdfstring{$V$}{V} Results}
\label{app:chi-square}

For each pairwise comparison declared in \S\ref{subsec:regime-annotation}, we form a $2 \times 4$ contingency table on regime counts (BOUNDARY, IP, DC, DM), excluding PARSE\_FAIL transitions, and compute Pearson's $\chi^2$ statistic, p-value, and bias-corrected Cram\'er's $V$ \citep{bergsma2013bias}. Tests are independent (no multiple-comparison correction); we interpret results descriptively rather than inferentially. Effect-size labels follow the convention: $V < 0.1$ negligible, $0.1 \le V < 0.3$ small, $0.3 \le V < 0.5$ medium, $V \ge 0.5$ large.

\paragraph{Axis 1: Cross-provider at matched benchmark difficulty (debate).} Table~\ref{tab:chi-axis1} shows that cross-provider differences in regime distribution are highly significant but with small effect sizes ($V \in [0.07, 0.18]$).

\begin{table*}[t]
\centering
\small
\begin{tabular}{llrrr}
\toprule
Cohort A & Cohort B & $\chi^2$ & p & $V$ \\
\midrule
gpt-4.1 MATH-h & Gemini MATH-h & 609.4 & $<10^{-4}$ & 0.182 \\
gpt-4.1 GPQA-D & Gemini GPQA-D & 398.5 & $<10^{-4}$ & 0.170 \\
gpt-4.1 MATH-h & nano MATH-h   & 184.6 & $<10^{-4}$ & 0.099 \\
gpt-4.1 MATH-h & gpt-oss-120b MATH-h & 305.4 & $<10^{-4}$ & 0.127 \\
Gemini MATH-h  & gpt-oss-120b MATH-h & 84.4  & $<10^{-4}$ & 0.066 \\
\bottomrule
\end{tabular}
\caption{Pairwise $\chi^2$ on regime distribution under debate, matched benchmark difficulty. All tests reject the null of identical distributions; Cram\'er's $V$ quantifies effect size.}
\label{tab:chi-axis1}
\end{table*}

\paragraph{Axis 2: Within-model across benchmarks (debate).} Table~\ref{tab:chi-axis2} contrasts gpt-4.1 and Gemini: gpt-4.1's regime distribution shifts substantially with benchmark difficulty ($V$ up to 0.26), while Gemini's barely shifts ($V = 0.09$, negligible).

\begin{table*}[t]
\centering
\small
\begin{tabular}{llrrr}
\toprule
Cohort A & Cohort B & $\chi^2$ & p & $V$ \\
\midrule
gpt-4.1 MATH-h & gpt-4.1 GPQA-D & 277.2 & $<10^{-4}$ & 0.130 \\
gpt-4.1 MATH-h & gpt-4.1 AIME   & 770.9 & $<10^{-4}$ & 0.260 \\
gpt-4.1 GPQA-D & gpt-4.1 AIME   & 450.5 & $<10^{-4}$ & 0.223 \\
Gemini MATH-h  & Gemini GPQA-D  & 141.8 & $<10^{-4}$ & 0.093 \\
\bottomrule
\end{tabular}
\caption{Pairwise $\chi^2$ on regime distribution, within model, across benchmarks (debate). The asymmetry between gpt-4.1 ($V$ up to 0.26) and Gemini ($V = 0.09$) quantifies the difference in benchmark-responsiveness: gpt-4.1's regime distribution responds to changing difficulty, Gemini's does not.}
\label{tab:chi-axis2}
\end{table*}

\paragraph{Axis 3: Cross-method within cohort (E1 vs E2).} Table~\ref{tab:chi-axis3} compares debate (E1) to self-correction (E2) within each model $\times$ benchmark cell. Most effect sizes are negligible, supporting the framework's claim that the detection threshold is a stable model/protocol-level regularity in our grid (\S\ref{subsec:detection-variation}). The exceptions are gpt-4.1 GPQA-D (small effect, $V = 0.121$) and Llama-1B GSM8K (medium effect, $V = 0.315$); the latter is consistent with Llama-1B's qualitatively different round dynamics under the two methods (\S\ref{subsec:dynamics}).

\begin{table*}[t]
\centering
\small
\begin{tabular}{llrrr}
\toprule
Model $\times$ Benchmark & Test & $\chi^2$ & p & $V$ \\
\midrule
gpt-4.1 / MATH-h    & E1 vs E2 & 18.4  & $0.0004$   & 0.039 \\
gpt-4.1 / GPQA-D    & E1 vs E2 & 114.4 & $<10^{-4}$ & 0.121 \\
Gemini / MATH-h     & E1 vs E2 & 2.5   & $0.4791$   & 0.000 \\
Gemini / GPQA-D     & E1 vs E2 & 43.2  & $<10^{-4}$ & 0.074 \\
nano / MATH-h       & E1 vs E2 & 13.4  & $0.0039$   & 0.032 \\
gpt-oss-120b / MATH-h & E1 vs E2 & 14.4 & $0.0024$  & 0.033 \\
Llama-1B / GSM8K    & E1 vs E2 & 387.3 & $<10^{-4}$ & 0.315 \\
\bottomrule
\end{tabular}
\caption{Pairwise $\chi^2$ on regime distribution, comparing debate (E1) against self-correction (E2) within each cohort. Most effect sizes are negligible---the regime mix is largely method-invariant within a model $\times$ benchmark cell. The Llama-1B exception is consistent with its qualitatively different round dynamics under the two methods.}
\label{tab:chi-axis3}
\end{table*}


\section{Annotation Protocol}
\label{app:annotation-protocol}

This appendix summarizes the protocol used for the two annotation studies reported in Appendix~\ref{app:annotation}: the rubric edge-case validation (30 cases, single annotator) and the reverse-mechanism taxonomy (50 cases, two annotators). Full verbatim guidelines and worked examples are released in the supplementary materials (\texttt{GUIDELINES.md}, \texttt{annotation\_examples.md}, \texttt{annotation\_dataset\_plan.md}).

\paragraph{Annotators.} Two research collaborators served as annotators (referred to as ``Annotator A'' and ``Annotator B'' in Appendix~\ref{app:annotation}). They are not crowdworkers; no compensation beyond standard research collaboration was involved. Annotators received instructions through the written guidelines documented in the supplementary materials, supplemented by informal communication.

\paragraph{Setup.} Annotators received CSV files with rows pre-populated with the agent transcript (problem statement, response excerpts, parsed answers, ground truth) and blank label columns. Annotators filled in label columns using a spreadsheet tool (Excel, Numbers, or Google Sheets). Approximate annotation time per task: 1--2 hours for the rubric edge-case validation (30 rows) and 3 hours for the reverse-mechanism taxonomy (50 rows).

\paragraph{Rubric edge-case validation (30 cases).} The annotator independently judged the detection label $D \in \{0, 1\}$ and the regime label (BOUNDARY, IP, DC, or DM) for each transition, then we compared against the deterministic rubric's output. The 30-transition sample was stratified across five edge-case types: 10 format-equivalent (answers differ by whitespace or LaTeX wrappers), 5 algebraic-equivalent (e.g., $12\pi$ vs $37.7$), 5 parse-failure (no extractable answer at one end), 5 hedge/multi-answer (response presents multiple candidates), and 5 random-baseline transitions drawn from the gpt-4.1 / MATH-hard / debate cohort.

\paragraph{Reverse-mechanism taxonomy (50 cases, two annotators).} Each of two annotators independently labeled all 50 reverse-direction events ($R_k \to R_{k+1}$ transitions where a correct upstream answer was revised to incorrect) using a four-way taxonomy:
\begin{itemize}
    \item \textbf{intrinsic\_self\_doubt}: agent revised the correct answer without external prompting (e.g., explicit re-evaluation phrases such as ``let me reconsider'').
    \item \textbf{specific\_peer\_argument\_convinced}: in debate, the agent flipped its answer after engaging with a specific peer argument.
    \item \textbf{recomputation\_slip}: agent re-derived the answer from scratch and arithmetic or algebra slipped during the redo.
    \item \textbf{other\_unclear}: none of the above clearly applies (e.g., extraction artifact, degenerate looping, format-only change).
\end{itemize}

\paragraph{Adjudication.} No re-annotation pass was performed after the initial annotation; the original labels are reported as-is. Inter-annotator agreement (Cohen's $\kappa$) on the reverse-mechanism taxonomy is reported in Appendix~\ref{app:annotation}; the dominant disagreement source (the boundary between \emph{intrinsic\_self\_doubt} and \emph{recomputation\_slip}) is documented there as a target for future taxonomy refinement.

\end{document}